\documentclass[aps,twocolumn,prl,amsmath,superscriptaddress,floatfix,showpacs,showkeys]{revtex4-1}
\usepackage{graphicx}
\usepackage{hyperref}

\newcommand{\ve}[1]{\mathbf{#1}}

\begin{document}

\title{The Role of Short-Range Order and Hyperuniformity in the Formation of Band Gaps in Disordered Photonic Materials}

\author{Luis S.\ Froufe-P\'erez}
\affiliation{Department of Physics, University of Fribourg, CH-1700 Fribourg, Switzerland}

\author{Michael Engel}
\affiliation{Department of Chemical Engineering, University of Michigan, Ann Arbor, Michigan 48109, USA}
\affiliation{Biointerfaces Institute, University of Michigan, Ann Arbor, Michigan 48109, USA}

\author{Pablo F.\ Damasceno}
\affiliation{Department of Chemical Engineering, University of Michigan, Ann Arbor, Michigan 48109, USA}
\affiliation{Biointerfaces Institute, University of Michigan, Ann Arbor, Michigan 48109, USA}

\author{Nicolas Muller}
\affiliation{Department of Physics, University of Fribourg, CH-1700 Fribourg, Switzerland}

\author{Jakub Haberko}
\affiliation{Faculty of Physics and Applied Computer Science, AGH University of Science and Technology,  al. Mickiewicza 30, 30-059 Krakow, Poland}

\author{Sharon C.\ Glotzer}
\affiliation{Department of Chemical Engineering, University of Michigan, Ann Arbor, Michigan 48109, USA}
\affiliation{Biointerfaces Institute, University of Michigan, Ann Arbor, Michigan 48109, USA}
\affiliation{Department of Materials Science and Engineering, University of Michigan, Ann Arbor, Michigan 48109, USA}

\author{Frank Scheffold}
\affiliation{Department of Physics, University of Fribourg, CH-1700 Fribourg, Switzerland}

\begin{abstract}
We study photonic band gap formation in two-dimensional high refractive index disordered materials where the dielectric structure is derived from packing disks in real and reciprocal space.
Numerical calculations of the photonic density of states demonstrate the presence of a band gap for all polarizations in both cases.
We find that the band gap width is controlled by the increase in positional correlation inducing short-range order and hyperuniformity concurrently.
Our findings suggest that the optimization of short-range order, in particular the tailoring of Bragg scattering at the isotropic Brillouin zone, are of key importance for designing disordered PBG materials.
\end{abstract}

\maketitle


Photonic band gap (PBG) materials exhibit frequency bands where the propagation of light is strictly prohibited.
Such materials are usually designed by arranging high refractive index dielectric material on a crystal lattice~\cite{Joannopoulos_book,Cefe_Adv_Mat_review_2003}.
The description of wave transport in a periodically repeating environment provides a clear physical mechanism for the emergence of PBGs, in analogy to common electronic semiconductors. 
It is also known that certain aperiodic dielectric structures, such as quasicrystals~\cite{Netti_Nature_2000,Steinhardt_PRL_2008,Edagawa_Sci_Tech_Adv_Mat_2014}, can display a full PBG.
Over the last decade disordered or amorphous photonic materials have gained growing attention~\cite{Zhang_PRB_2001,Rockstuhl_Opt_Lett_2006,Jian_Zi_Adv_Mat_2013,Scheffold_PRL_2004,markovs2005intensity,Scheffold_APL_2007,Galisteo_adv_mat_2011,Segev_PRL_2011,izrailev2012anomalous,Wiersma_Nat_Phot_2013,Gora_Wiersma_PRL_2014,Scheffold_OPEX_2013,Muller_Scheffold_Adv_Opt_Mat_2014}.
This trend is motivated by the many disordered photonic materials found in nature that reveal fascinating structural color effects in plants, insects, and mammals~\cite{Jin_Tong_RCS_Adv_2013}.
At the same time, fabricating perfect crystalline structures with photonic properties at optical wavelengths has proven to be more difficult than initially anticipated~\cite{lourtioz2004photonic}. It has been argued that disordered PBG materials should be less sensitive to fabrication errors or defects and thus promise a more robust design platform~\cite{Wiersma_Nat_Phot_2013}.
Moreover PBGs in disordered dielectrics are isotropic, which could make it easier to achieve a full PBG while at the same time offering better performance in wave guiding, design of non-iridescent stable pigments, and display applications~\cite{man2013isotropic,Takeoka2012,Vinothan_Ang_Chem_2014,Dyachenko2014}. 
\newline \indent Yet, until recently, direct evidence for the existence of full PBGs in disordered photonic materials had been scarce and the fabrication principles and physical-optical mechanism leading to PBG formation remained obscure.
Although the importance of appropriate short-range order for the development of PBGs in disordered photonic materials was discovered early on~\cite{Zhang_PRB_2001,Rockstuhl_Opt_Lett_2006,Jian_Zi_Adv_Mat_2013,Scheffold_PRL_2004}, a strategy to maximize the PBG width was lacking.

In 2009 Florescu and coworkers~\cite{Florescu_PNAS_2009} proposed a new approach for the design of disordered PBG materials that has attracted widespread attention.
They introduced the concept of hyperuniformity for photonic structures, which enforces a certain type of short-range order.
In particular, so-called stealthy hyperuniform (SHU) disordered patterns were reported to be fully transparent to incident long-wavelength radiation~\cite{Torquato_J_App_Phys_2008,Leseur2015} and lead to strong isotropic PBGs at shorter wavelengths~\cite{Florescu_PNAS_2009}.
Other types of correlated disorder, such as those generated by the random-sequential absorption model, were claimed to be inferior because they do not induce the formation of PBGs for both polarizations simultaneously~\cite{Florescu_PNAS_2009}.
To the contrary, independent numerical work showed that the amorphous diamond structure~\cite{Edagawa2008} and three-dimensional networks derived from systems of densely packed spheres~\cite{Cao_PRA_2011} exhibit a full PBG in the absence of stealthiness.

Here we aim to disentangle the role of short-range order and hyperuniformity in producing PBGs.
To establish a comparison between different disordered systems, we compare, using otherwise identical design protocols matching those studied in Ref.~\cite{Florescu_PNAS_2009}, the photonic properties of systems with varying degrees and types of correlations: (i)~collections of rods for transverse-magnetic (TM) polarization, (ii)~trivalent networks (connected walls) for transverse-electric (TE) polarization, and (iii)~decorated trivalent networks for both polarizations.
Our results suggest that stealthy hyperuniformity and disk packing are equivalent strategies for the design of two-dimensional disordered photonic materials.


\emph{Pattern generation.}---Hyperuniformity corresponds to the suppression of long-range density fluctuations and can be expressed as a condition for the structure factor, $S(k\rightarrow0)=0$.
Stealthiness is even more restrictive and requires the structure factor to vanish over a finite range, $S(k)=0$ for $k<K$~\cite{Torquato_J_App_Phys_2008}.
Importantly, two-dimensional SHU patterns can be regarded as a reciprocal-space counterpart of packing hard disks (HDs)~\cite{Torquato2015} of radius $R$, which fulfill the condition $g(r<R)=0$ for the radial distribution function.
In both cases the excluded zone is determined by a dimensionless parameter, the packing fraction $\phi$ in real space and stealthiness $\chi$ in reciprocal space~\cite{Torquato_J_App_Phys_2008}.
We interpret these parameters as measures for the amount of correlation in the system restricting randomness.
Increasing $\phi$ or $\chi$ leads to pronounced peaks in $g(r)$ and $S(k)$, indicating the development of short-range positional order.
Upon further increase of $\phi$ or $\chi$, entropy-driven crystallization sets in eventually.
Moreover, hyperuniformity is recovered asymptotically, $S(k) \to 0$ for $k\to 0$, when disks are packed into a maximally jammed configuration via compression while avoiding crystallization~\cite{Torquato_PRL_2005,Berthier_arXiv_Apr_2015}.

We generate disordered point patterns with different levels of positional correlation as measured by $\phi$ and $\chi$ using computer simulation~\cite{Supplemental}.
Disordered HD patterns are equilibrated fluids.
SHU patterns are obtained using a pair potential derived from the potential energy, $E=\sum_{|\ve{k}|\leq K}S(\ve{k})$, where the discreteness of the sum is a consequence of the periodic simulation box used.
We employ a simulated annealing relaxation scheme to find disordered SHU patterns with $S(k)<10^{-6}$ for $k<K(\chi)$.
Selecting an area $A$ in real space for a given number of points $N$ sets the number density $\rho=N/A$ and defines a characteristic length scale $a=\rho^{-1/2}$, comparable to a typical distance between the points.

\begin{figure}
\includegraphics[width=0.73\columnwidth]{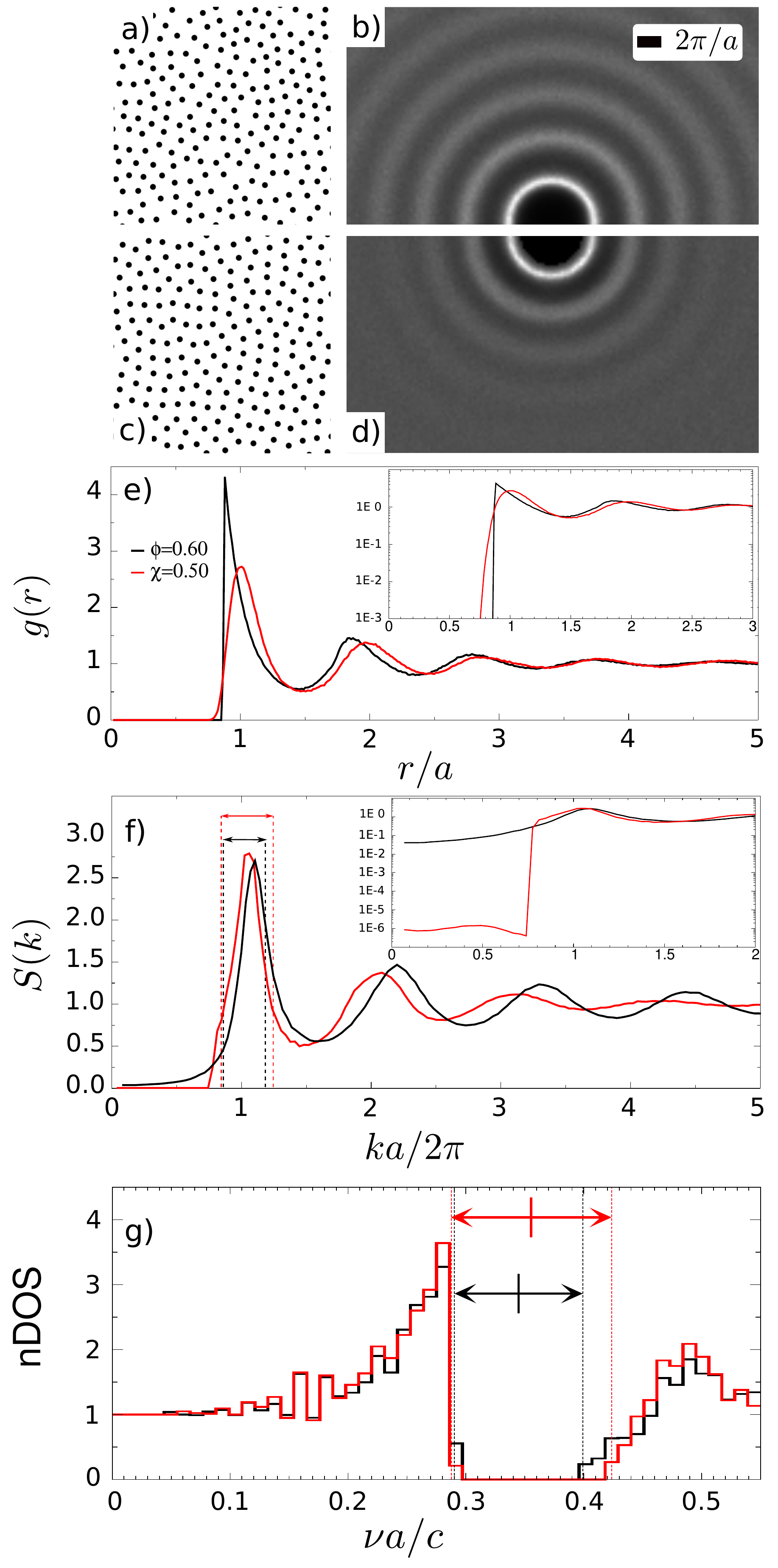}
\caption{\label{fig:1}
Structural order in (a,b,e,f)~HD patterns at $\phi=0.60$ and (c-f)~SHU patterns at $\chi=0.50$.
We compare (a,c)~typical point patterns in real space, (b,d)~structure factors $S(\ve{k})$, (e)~radial distribution functions $g(r)$, and (f)~angle-averaged structure factors $S(k)$.
The two-dimensional plots in (b,d) confirm that the distributions are isotropic.
The inset in (e) shows that a correlation gap is present in HD and SHU packings for $r<a$.
The inset in (f) demonstrates that long-wave length fluctuations $2\pi/k>2\pi/K$ are strongly suppressed (up to numerical noise) in the SHU pattern. All data is obtained by taking an average over 1000 independent patterns with $N=200$ points each. Vertical lines indicate the position of the band edges in reciprocal space for TM polarization. 
(g) Photonic properties of a dielectric structure composed of a network of silicon rods placed at the position of the points as shown in (a,c). nDOS as a function of the normalized frequency $\nu a/c$ in TM polarization for the HD system at $\phi=0.60$ and the SHU system at $\chi=0.50$.
Here, $c$ is the speed of light in vacuum.
}
\end{figure}

The patterns calculated with our algorithms quantitatively reproduce previously reported~\cite{Torquato2015} statistics and correlation functions.
Patterns below the critical parameters $\phi,\chi\ge0.70$, in which quasi-long-range order gradually appears~\cite{Werner_PRL_2011,Werner_PRE_2013,Torquato2015}, already have significant short-range order.
Fig.~\ref{fig:1}(a-f) shows representative HD and SHU point patterns at $\phi=0.60$ and $\chi=0.50$, respectively.
These parameter values were chosen as high as possible while retaining the amorphous structure and matching the $S(k)$ peaks as closely as possible.
A close similarity of the local short-range positional order in both patterns is visually apparent.
We now investigate the effect of this similarity on photonic properties. 

\emph{Photonic density of states}---We study numerically the photonic properties of a dielectric system composed of silicon (dielectric constant or permittivity $\epsilon=11.6$) rods and walls derived from HD and SHU seed patterns.
For the first protocol we place cylindrical rods (in three dimensions, disks in the plane) at the points of the seed pattern.
We use a fixed rod radius $r/a=0.189$, which leads to an area filling fraction $N\pi r^2/A=\pi (r/a)^2=0.112$ of the two-dimensional plane with silicon, independent of $\phi$ and $\chi$ provided  no two rods overlap.
The no-overlap condition is almost exactly met in the relevant regime $\phi,\chi\ge 0.2$.
We use the supercell method~\cite{Joannopoulos_book} implemented in the open source code MIT Photonic Bands~\cite{Johnson2001_mpb} to obtain the normalized photonic density of states (nDOS).
A finite sample (the supercell) is repeated periodically and the band structure calculated by following the path $\Gamma\rightarrow M\rightarrow X\rightarrow\Gamma$ in reciprocal space.
We accumulate a histogram of eigenfrequencies and compare it to the corresponding one for a homogeneous medium with a refractive index that best matches the first band (low energy) of both the structured and the homogeneous medium.
In the next step, the effective refractive index of the structure and the width and position of the PBGs are obtained by direct analysis of the band structure~\footnote{Because we determine the PBG width directly from the band structure, the precision is higher than suggested from the bin width of the histogram, \emph{e.g.}\ in Fig.~\ref{fig:1}(g)}.

The nDOS for the HD system with $\phi=0.60$ and for the SHU system with $\chi=0.50$ [Fig.~\ref{fig:1}(g)] are almost identical in TM polarization and resemble a typical nDOS of a photonic crystal.
Apparently the structural similarity of the seed patterns translates into a corresponding similarity of the photonic band structure.
In fact, the position and width of the PBG covers exactly the peak region of $S(k)$, Fig.~\ref{fig:1}(f), corresponding to Bragg scattering at the isotropic Brillouin zone.
\begin{figure}
\includegraphics[width=0.8\columnwidth]{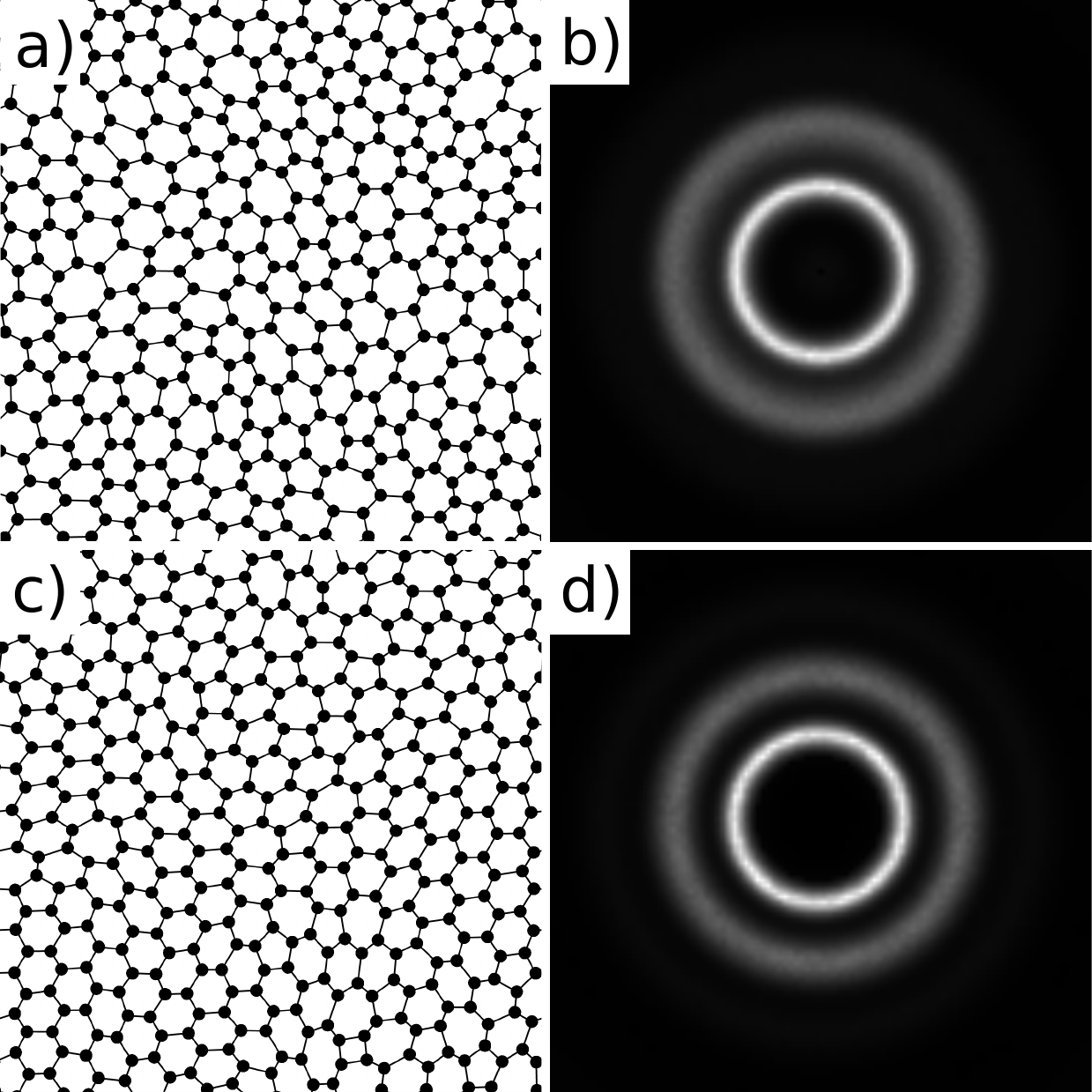}
\caption{\label{fig:2}
Decorated network structure derived from (a) a HD seed pattern with $N=200$, $\phi=0.60$ and (c) a SHU seed pattern with $N=200$, $\chi=0.50$. 
(b) and (d) show the average of the square amplitude of the corresponding Fourier transforms taken over 1000 network realizations.
}
\end{figure}

It is known~\cite{Joannopoulos_JOSAB_1993}  that the placement of rods is a good ansatz to obtain a PBG in TM polarization but does not lead to the opening of a PBG in TE polarization, even for perfect crystals. Instead, in the second protocol, we perform the Delaunay triangulation of the seed pattern and connect the barycenters of neighboring triangles by walls (in three dimensions, bonds in the plane) to form a trivalent network~\footnote{In the process of triangulation and barycenter linking the number of network nodes (N) of the seed pattern doubles.}.
This protocol enforces a uniform structure with three bonds at each node, favouring the opening of a gap due to the local topology \cite{Florescu_PNAS_2009,weaire1971existence,weaire1971electronic}.
We use a fixed wall thickness of $w/a=0.288$.
Finally, to enforce PBGs in both polarizations simultaneously, we combine the two protocols and form a decorated network consisting of rods (radius $r/a=0.2275$) and Delaunay walls (thickness $w/a=0.0593$) [Fig.~\ref{fig:2}]. All geometric parameters are optimized to yield a maximally wide PBG, see also supplemental material ~\cite{Supplemental}. The results for the HD system and the SHU system in TM, TE polarization as well as in both polarizations simultaneously are equally similar~\cite{Supplemental}.

\begin{figure}
\includegraphics[width=.8\columnwidth]{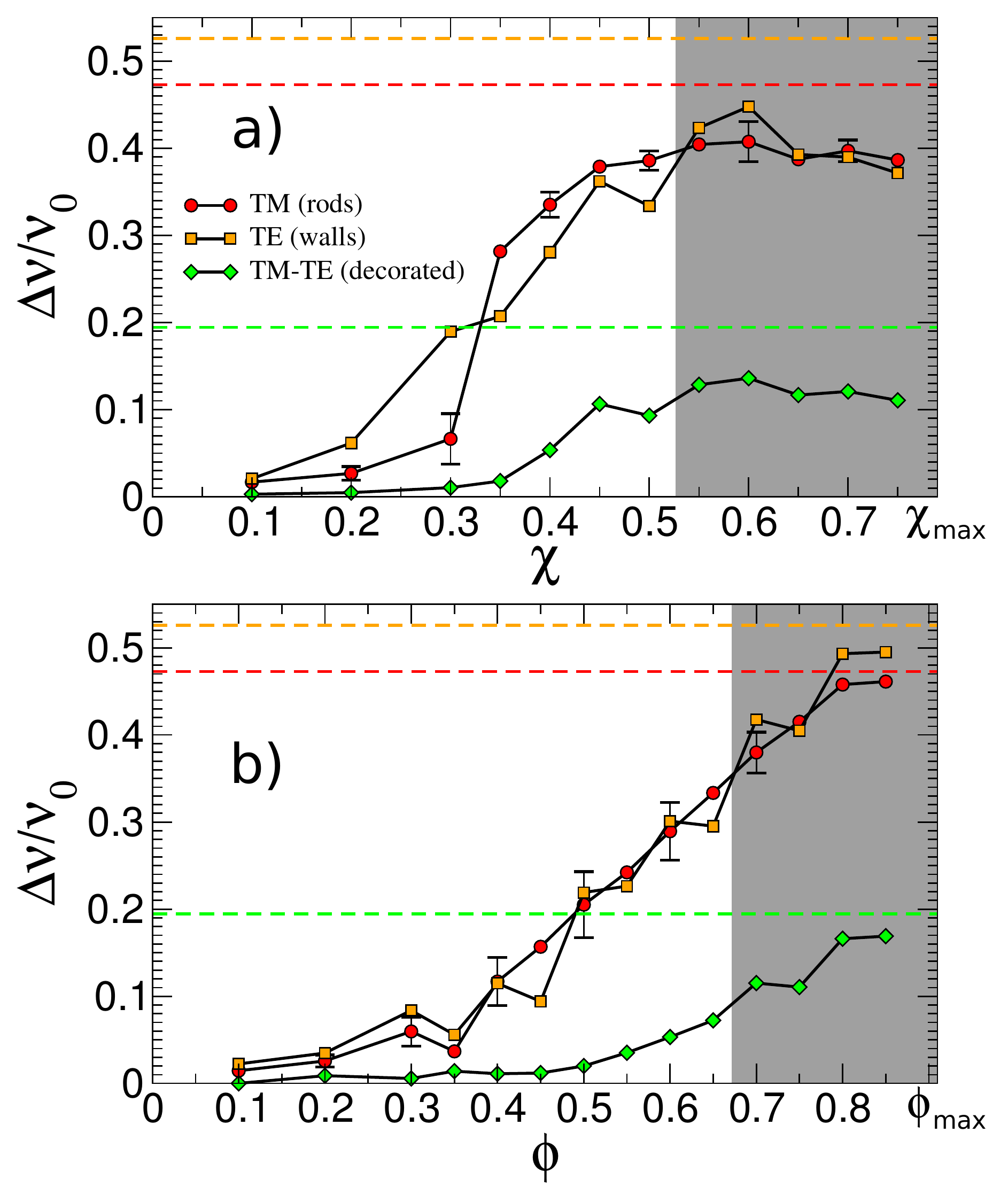}
\caption{\label{fig:3}
Formation of PBGs for TM modes, TE modes and for both polarizations simultaneously.
The normalized PBG width is plotted as a function of the parameters $\chi$ and $\phi$ for (a)~SHU and (b)~HD seed patterns up to a maximal value of $\chi_\text{max}=0.79$ and $\phi_\text{max}=0.91$. The shaded area marks the appearance of first signs of peaks in reciprocal space indicating the presence of quasi-long-range crystalline order.
The dashed lines indicate the normalized PBG widths for otherwise identical structural designs when the points are arranged according to a hexagonal lattice seed pattern.}
\end{figure}

Next, we analyze the normalized PBG width for TM, TE and both polarizations simultaneously as a function of stealthiness $\chi$ and packing density $\phi$. As shown in Fig.~\ref{fig:3}, the PBG is initially narrow but widens with increasing $\chi$ and $\phi$.
Interestingly, the central frequency $\nu_0$ of the PBG is almost unchanged at $\nu_0 a/c\simeq0.35$.
$\nu_0$ also coincides with the PBG center of a hexagonal lattice with identical scatterer geometry and area filling fraction.
For comparison we include in Fig.~\ref{fig:3} results obtained when the seed pattern is the hexagonal lattice.
In this case, the PBG is even wider, which suggests that the amorphous patterns attempt to asymptotically reach the crystal values but cannot exceed them -- at least not for the two-dimensional structures considered here. 
The joint PBG for TM and TE polarization is generally narrower than the PBG for TM or TE polarization alone and typically reaches only a width of approximately $13\%$ for amorphous patterns and $19\%$ for the hexagonal lattice.

\begin{figure}
\includegraphics[width=1\columnwidth]{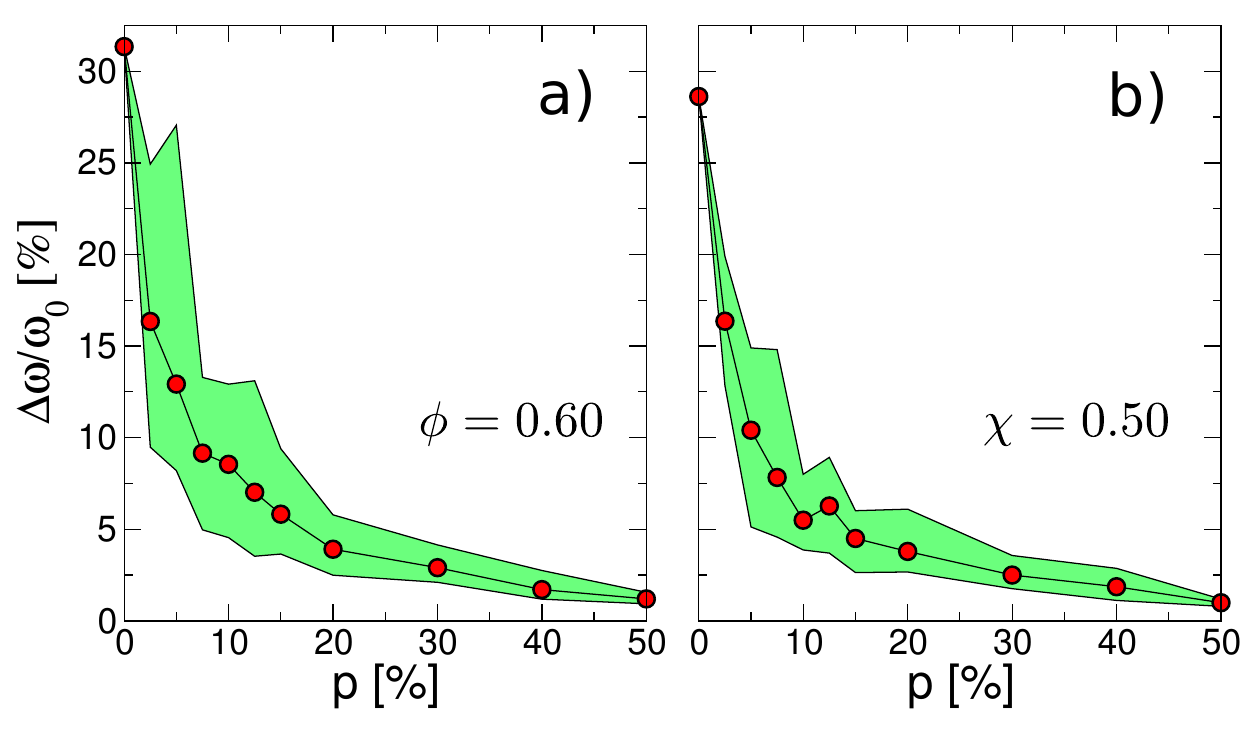}
\caption{\label{fig:4}
Robustness of the band gap for a HD structure at $\phi=0.6$~(a) and a SHU structure at $\chi=0.5$~(b).
The plot shows the averaged normalized PBG width for networks of dielectric walls where a percentage $p$ of links is removed randomly.
The shaded area represent the smallest and largest gap within the ensemble of ten configurations studied. The calculation was performed in TE polarization.}
\end{figure}

\emph{Robustness of the photonic band gap}---Finally, we address briefly the influence of imperfections on the width of the PBG.
Imperfections are unavoidable in the experimental realization of photonic materials and the robustness of the PBG is an important design parameter.
For simplicity we restrict our analysis to the case of random link removal in the network of walls and for TE polarization, following the procedure described in ~\cite{florescu2010effects}.
A percentage $p$ of links is removed randomly and for each value of $p$ we calculate the nDOS for ten different structures.
As shown in Fig.~\ref{fig:4}, we find that the removal of links gradually reduces the width of the gap. Our results indicate that, within statistical error, the results are the same for the SHU and the HD structure. Moreover, our results are similar to those obtained for the corresponding hexagonal lattices of dielectric walls~\cite{florescu2010effects}, see also supplemental material ~\cite{Supplemental}.

\emph{Discussion and conclusion.}---Our findings demonstrate that disordered packings in real and reciprocal space are equally suitable for generating isotropic PBG materials in two dimensions.
SHU patterns restrict the accessible phase space in a different way, but are not found to be more efficient in opening wider PBGs than HD patterns.
The parameters $\phi$ and $\chi$ acquire their meaning due to the presence of a maximum value, \emph{i.e.}\ a value where the accessible phase space volume is zero, the system is fully constrained, and dynamics disappears.
In the case of HDs this value is the packing density of the densest packing of disks in the hexagonal lattice, $\phi_\text{max}=\pi/\sqrt{12}\simeq0.91$.
For SHU patterns the maximum value is obtained once the excluded zone in reciprocal space touches the Bragg peaks.
Interestingly, and for reasons that are unclear, our results suggest that SHU patterns converge towards the square lattice~\cite{Supplemental} after the appearance of intermediate stacked-slider phases with local hexagonal order~\cite{Zhang2015}, and thus $\chi_\text{max}=\pi/4\simeq0.79$.
Since the phase space restrictions that are imposed by $\chi$ and $\phi$ are highly non-linear, no simple linear relationship exists between them well below their maximum value.

The characteristic length scale for the development of a PBG, the so-called Bragg length $l_B$, is typically on the order of the characteristic structural length scale $a$.
Suppression of scattering at wave numbers smaller than $2\pi/l_B$, as targeted by hyperuniformity, should therefore play a marginal role in the formation of the PBG.
Indeed, we believe that the emergence of a PBG is a side-effect of hyperuniformity. Increasing $\chi$ prevents scattering for $k<K$ and the intensity piles up, due to a sum rule, just above $K$~\footnote{this point was first raised in discussions with Chris Henley, Cornell University, private communication (2014).}.
Suppression of scattering at small wave numbers and hyperuniformity (strictly or in approximation) are then natural consequences of the development of short-range order and vice versa. 

In all cases studied the maximum bandwidth approaches the crystal values only asymptotically.
The influence of defects is local and it affects the band gap most whenever such defects percolate as shown in \cite{florescu2010effects}.
Moreover in \cite{Cao_PRA_2010} it was found that the bandgap of polycrystalline photonic structures remains nearly unchanged from that of the perfect crystalline structure as long as the crystal domains are larger than a few times the Bragg length $l_B$.
These observations, taken together, suggest that the band gap formation is determined by local properties.
Optimization of short-range order, in particular the tailoring of Bragg scattering at the isotropic Brillouin zone, and the appropriate topology are expected to be the key aspects for enforcing a photonic band gap in dielectric materials.   

This research was supported by the National Science Foundation, Division of Materials Research Award No.~DMR 1120923 (M.E., P.F.D., S.C.G.).
S.C.G.\ acknowledges partial support by a Simons Investigator award from the Simons Foundation.
This research was supported by the Swiss National Science Foundation through the National Centre of Competence in Research \emph{Bio-Inspired Materials} and through project number 149867 (L.S.F.P, N.M., F.S.).
N.M. acknowledges support by the Luxembourg National Research fund (project No. 3093332). L.S.F.P., N.M., and F.S.\ would like to thank Kevin Vynck for interesting discussions.

\end{document}